\documentclass[conference]{IEEEtran}
\IEEEoverridecommandlockouts
\usepackage{amsmath,amssymb,amsfonts}
\usepackage{algorithmic}
\usepackage{graphicx}
\usepackage{textcomp}
\def\BibTeX{{\rm B\kern-.05em{\sc i\kern-.025em b}\kern-.08em
    T\kern-.1667em\lower.7ex\hbox{E}\kern-.125emX}}

\usepackage[url,hyperrefblack,notheorems,IEEEtran]{research17}
\usepackage{lipsum} 
 \usepackage{balance}
\usepackage{booktabs} 
\usepackage{enumitem}
\usepackage{mathtools}
\usepackage[lined,boxed,commentsnumbered,linesnumbered,ruled]{algorithm2e}
\usepackage{tikz} 
\usetikzlibrary{shapes, patterns,decorations.text, decorations.pathreplacing}

\newtheorem{theorem}{\mytheoremname}
\newtheorem{definition}{\mydefinitionname}
\newtheorem{proposition}{\mypropositionname}
\newtheorem{corollary}{\mycorollaryname}
\newtheorem{example}{\myexamplename}

\usepackage[capitalize]{cleveref}
\crefname{equation}{\unskip}{\unskip}
\crefname{claim}{Claim}{Claims} 
\usepackage{array}
\usepackage{multirow}
\newcolumntype{C}[1]{>{\centering\arraybackslash}p{#1}}
\allowdisplaybreaks

\renewcommand{\vect}[1]{\vectg{#1}} 
\renewcommand{\mat}[1]{\bm{#1}} 
\renewcommand{\vmat}[1]{\bm{#1}} 
\newcommand{\code}[1]{\mathcal{#1}} 
\newcommand{\Nat}[1]{\mathbb{N}_{#1}} 
\newcommand*{\Scale}[2][4]{\scalebox{#1}{\ensuremath{#2}}} 
\def\rot#1{\rotatebox{90}{#1}} 
\makeatletter
\renewcommand*\env@matrix[1][*\c@MaxMatrixCols c]{%
  \hskip -\arraycolsep
  \let\@ifnextchar\new@ifnextchar
  \array{#1}}
\makeatother

\renewcommand{\HH}{\mathop{}\!\mathsf{H}}  
\newcommand{\HP}[1]{\HH\left(#1\right)} 
\newcommand{\eHP}[1]{\HH(#1)}

\newcommand{\HPcond}[2]{\HH\left(#1 \kern0.1em\middle|\kern0.1em #2\right)}
\newcommand{\eHPcond}[2]{\HH(#1 \kern0.1em|\kern0.1em #2)} 
\newcommand{\bigHPcond}[2]{\HH\bigl(#1 \kern-0.1em \bigm| \kern-0.1em#2\bigr)}
\newcommand{\BigHPcond}[2]{\HH\Bigl(#1 \kern-0.1em \Bigm| \kern-0.1em#2\Bigr)}
\renewcommand{\II}{\mathop{}\!\mathsf{I}}  
\newcommand{\MI}[2]{\II\left(#1 \kern0.1em{;}\kern0.1em #2\right)} 
\newcommand{\eMI}[2]{\II(#1 \kern0.1em{;}\kern0.1em #2)} 
\newcommand{\bigMI}[2]{\II\bigl(#1 \kern0.1em{;}\kern0.1em #2\bigr)}
\newcommand{\BigMI}[2]{\II\Bigl(#1 \kern0.1em{;}\kern0.1em #2\Bigr)}
\newcommand{\MIcond}[3]{\II\left(#1 \kern0.1em{;}\kern0.1em #2 \kern0.1em\middle|\kern0.1em #3\right)}
\newcommand{\eMIcond}[3]{\II(#1 \kern0.1em{;}\kern0.1em #2 \kern0.1em|\kern0.1em #3)} 
\newcommand{\bigMIcond}[3]{\II\bigl(#1 \kern0.1em{;}\kern0.1em #2 \kern-0.1em \bigm| \kern-0.1em#3\bigr)}
\newcommand{\BigMIcond}[3]{\II\Bigl(#1 \kern0.1em{;}\kern0.1em #2 \kern-0.1em \Bigm| \kern-0.1em#3\Bigr)}
\renewcommand{\r}{\color{black}} 
\renewcommand{\b}{\color{black}} 

\begin{document}

\title{Asymmetry Helps: Improved Private Information Retrieval Protocols for  Distributed Storage\thanks{This work
    was partially funded by the Research Council of Norway (grant 240985/F20) and the Swedish Research Council (grant
    \#2016-04253).}}


\author{\IEEEauthorblockN{Hsuan-Yin Lin\IEEEauthorrefmark{2}, Siddhartha Kumar\IEEEauthorrefmark{2}, Eirik
    Rosnes\IEEEauthorrefmark{2}, and Alexandre Graell i Amat\IEEEauthorrefmark{3}}
  \IEEEauthorblockA{\IEEEauthorrefmark{2}Simula UiB, N--5020 Bergen, Norway}
  \IEEEauthorblockA{\IEEEauthorrefmark{3}Department of Electrical Engineering, Chalmers University of Technology,
    SE--41296 Gothenburg, Sweden}}

\maketitle

\begin{abstract}
  We consider private information retrieval (PIR) for distributed storage systems (DSSs) with noncolluding nodes where
  data is stored using a non maximum distance separable (MDS) linear code.  It was recently shown that if data is stored
  using a particular class of non-MDS linear codes, the \emph{MDS-PIR capacity}, i.e., the maximum possible PIR rate for
  MDS-coded DSSs, can be achieved.  For this class of codes, we prove that the PIR capacity is indeed equal to the
  MDS-PIR capacity, giving the first family of non-MDS  codes for which the PIR capacity is known.
  For other codes, we provide asymmetric PIR protocols that achieve a strictly larger PIR rate compared to
  existing symmetric PIR protocols.
  %
  %
\end{abstract}

\section{Introduction}

The concept of private information retrieval (PIR) was first introduced by Chor \emph{et al.}
\cite{ChorGoldreichKushilevitzSudan95_1}. A PIR protocol allows a user to privately retrieve an arbitrary data item
stored in multiple servers (referred to as nodes in the sequel) without disclosing any information of which item is requested 
to the nodes. The efficiency of a PIR protocol is measured in terms of the total communication cost between the user and
the nodes, which is equal to the sum of the upload and download costs. In distributed storage systems (DSSs), data is
encoded by an $[n,k]$ linear code and then stored on $n$ nodes in a distributed manner. Such DSSs are referred to as
coded DSSs \cite{ShahRashmiRamchandran14_1,ChanHoYamamoto15_1}.

One of the primary aims in PIR is the design of efficient PIR protocols from an information-theoretic perspective. Since
the upload cost does not scale with the file size, the download cost dominates the total communication cost
\cite{ChanHoYamamoto15_1,TajeddineElRouayheb16_1}. Thus, the efficiency of a PIR protocol is commonly measured by the
amount of information retrieved per downloaded symbol, referred to as the PIR rate. Sun and Jafar derived the
maximum achievable PIR rate, the so-called \emph{PIR capacity}, for the case of DSSs with replicated data
\cite{SunJafar17_1, SunJafar18_2}. In the case where the data stored is encoded by an MDS storage code (the so-called
\emph{MDS-coded DSS}) and no nodes collude, a closed-form expression for the PIR capacity, referred to as the
\emph{MDS-PIR capacity}, was derived in \cite{BanawanUlukus18_1}.


In the earlier work \cite{KumarRosnesGraellAmat17_1,KumarLinRosnesGraellAmat17_1sub,LinKumarRosnesGraellAmat18_1}, the
authors focused on the properties of non-MDS storage codes in order to achieve the MDS-PIR capacity. In particular, in
\cite{LinKumarRosnesGraellAmat18_1,KumarLinRosnesGraellAmat17_1sub} it was shown that the MDS-PIR capacity can be
achieved for a special class of non-MDS linear codes, which, with some abuse of language, we refer to as \emph{MDS-PIR
  capacity-achieving} codes (there might exist other codes outside of this class that achieve the MDS-PIR capacity). 
However, it is still unknown whether the
MDS-PIR capacity is the best possible PIR rate that can be achieved for an arbitrarily coded DSS.  In particular, an
expression for the PIR capacity for coded DSSs with arbitrary linear storage codes is still missing.

In this paper, we consider the noncolluding case and first prove that the PIR capacity of coded DSSs that use the class of MDS-PIR capacity-achieving codes
introduced in \cite{KumarLinRosnesGraellAmat17_1sub} is equal to the MDS-PIR capacity.  We then address the fundamental
question of what is the maximum achievable PIR rate for an arbitrarily coded DSS. To this purpose, we mainly consider
non-MDS-PIR capacity-achieving codes. Most of the earlier works focus on designing symmetric PIR protocols and it was
shown in \cite{SunJafar17_1,SunJafar17_2,BanawanUlukus18_1} that any PIR scheme can be made symmetric for MDS-coded
DSSs. However, this is in general not the case for non-MDS codes. Specifically, we propose an \emph{asymmetric} PIR
protocol, Protocol~A, that allows asymmetry in the responses from the storage nodes. For non-MDS-PIR capacity-achieving
codes, Protocol~A achieves improved PIR rates compared to the PIR rates of existing symmetric PIR protocols.
Furthermore, we present an asymmetric PIR protocol,  Protocol~B, that applies to non-MDS-PIR capacity-achieving
codes that can be written as a direct sum of MDS-PIR capacity-achieving codes. Finally, we give an example showing that
it is possible to construct an improved (compared to Protocol~A) asymmetric PIR protocol. The protocol is
code-dependent and strongly relies on finding \emph{good} punctured MDS-PIR capacity-achieving subcodes of the
non-MDS-PIR capacity-achieving code. 

\section{Preliminaries and System Model}
\label{sec:system-model}

\subsection{Notation and Definitions}

We denote by $\Nat{}$ the set of all positive integers and define $\Nat{a}\eqdef\{1,2,\ldots,a\}$. Vectors are denoted by
lower case bold letters, matrices by upper case bold letters, and sets by calligraphic upper case letters, e.g.,
$\vect{x}$, $\mat{X}$, and $\set{X}$ denote a vector, a matrix, and a set, respectively. In addition, $\comp{\set{X}}$
denotes the complement of a set $\set{X}$ in a universe set. 
The fonts of random and deterministic quantities are not distinguished typographically since it should be
clear from the context.
We denote a submatrix of $\mat{X}$ that is restricted in columns by the set $\set{I}$
by $\mat{X}|_{\set{I}}$. The function $\mathsf{LCM}(n_1,n_2,\ldots,n_a)$ computes the lowest common multiple of $a$
positive integers $n_1,n_2,\ldots,n_a$. The function $\HP{\cdot}$ represents the entropy of its argument and
$\MI{\cdot}{\cdot}$ denotes the mutual information of the first argument with respect to the second
argument.
$\trans{(\cdot)}$ denotes the transpose of its argument. We use the customary code parameters $[n,k]$
to denote a code $\code{C}$ over the finite field $\GF(q)$ of blocklength $n$ and dimension $k$.
A generator matrix of $\code{C}$ is denoted by $\mat{G}^{\code{C}}$, while $\code{C}^{\mat{G}}$ represents the
corresponding code generated by $\mat{G}$. The function $\chi(\vect{x})$ denotes the support of a vector
$\vect{x}$,
while the support of a code $\code{C}$ is defined as the set of coordinates where not all codewords are zero. A
set of coordinates of $\code{C}$, $\set{I}\subseteq\Nat{n}$, of size $k$ is said to be an \emph{information set} if and
only if $\mat{G}^\code{C}|_\set{I}$ is invertible. The $s$-th generalized Hamming weight of an $[n,k]$ code $\code{C}$,
denoted by $d_s^{\code{C}}$, $s\in \Nat{k}$, is defined as the cardinality of the smallest support of an $s$-dimensional
subcode of $\code{C}$.





\subsection{System Model}
\label{sec:system-model}


We consider a DSS that stores $f$ files $\mat{X}^{(1)},\ldots,\mat{X}^{(f)}$, where each file
$\mat{X}^{(m)}=(x_{i,l}^{(m)})$, $m\in\Nat{f}$, can be seen as a $\beta\times k$ matrix over $\GF(q)$ with
$\beta,k \in\Nat{}$. 
Each file is encoded using a linear code as follows. Let
$\vect{x}^{(m)}_i=\bigl(x^{(m)}_{i,1},\ldots,x^{(m)}_{i,k}\bigr)$, $i\in\Nat{\beta}$, be a message vector corresponding
to the $i$-th row of $\mat{X}^{(m)}$. Each $\vect{x}^{(m)}_i$ is encoded by an $[n,k]$ code $\code{C}$ over $\GF(q)$
into a length-$n$ codeword $\vect{c}^{(m)}_i=\bigl(c^{(m)}_{i,1},\ldots,c^{(m)}_{i,n}\bigr)$. The $\beta f$ generated
codewords $\vect{c}_i^{(m)}$ are then arranged in the array
$\mat{C}=\trans{\bigl(\trans{(\mat{C}^{(1)})}|\ldots|\trans{(\mat{C}^{(f)})}\bigr)}$ of dimensions $\beta f \times n$,
where $\mat{C}^{(m)}=\trans{\bigl(\trans{(\vect{c}^{(m)}_1)}|\ldots|\trans{(\vect{c}^{(m)}_{\beta})}\bigr)}$. The code
symbols $c_{1,l}^{(m)},\ldots,c_{\beta,l}^{(m)}$, $m\in\Nat{f}$, for all $f$ files are stored on the $l$-th storage
node, $l\in\Nat{n}$.

\subsection{Privacy Model}
\label{sec:privacy}


To retrieve file $\mat{X}^{(m)}$ from the DSS, the user sends a random query $Q_l^{(m)}$ to the $l$-th node for all
$l\in\Nat{n}$. In response to the received query, node $l$ sends the response $A^{(m)}_l$ back to the user. $A^{(m)}_l$
is a deterministic function of $Q_l^{(m)}$ and the code symbols stored in the node.
\begin{definition}
  \label{Def:perfect-PIR}
  Consider a DSS with $n$ noncolluding nodes storing $f$ files. A user who wishes to retrieve the $m$-th file sends the
  queries {\b{$Q^{(m)}_l$}}, $l\in\Nat{n}$, to the storage nodes, which return the responses $A^{(m)}_l$. This scheme
  achieves perfect information-theoretic PIR if and only if
  \begin{IEEEeqnarray}{rCl}
    \IEEEeqnarraymulticol{3}{l}{%
      \text{Privacy:} }\nonumber\\*\quad%
    && \bigMI{m}{Q^{(m)}_l,A^{(m)}_l,\vmat{X}^{(1)},\ldots,\vmat{X}^{(f)}}=0,\,\forall\,l\in\Nat{n},
    \IEEEyesnumber\IEEEyessubnumber\IEEEeqnarraynumspace\label{eq:privacy}
    \\
    \IEEEeqnarraymulticol{3}{l}{%
      \text{Recovery:} }\nonumber\\*\quad%
    && \bigHPcond{\vmat{X}^{(m)}}{A^{(m)}_1,\ldots,A^{(m)}_n,Q^{(m)}_1,\ldots,Q^{(m)}_n}=0.
    \IEEEyessubnumber\label{eq:recovery}
  \end{IEEEeqnarray}
\end{definition}




\subsection{PIR Rate and Capacity}
\begin{definition}
  \label{def:def_PIRrate}
  The PIR rate of a PIR protocol, denoted by $\const{R}$, is the amount of information retrieved per downloaded symbol,
  i.e., $\const{R}\eqdef\frac{\beta k}{\const{D}}$, where $\const{D}$ is the total number of downloaded symbols for the
  retrieval of a single file.
\end{definition}

We will write $\const{R}(\mathcal C)$ to highlight that the PIR rate depends on the underlying storage code
$\code{C}$.
%
It was shown in \cite{BanawanUlukus18_1} that for the noncolluding case and for a given number of
files $f$ stored using an $[n,k]$ MDS code, the MDS-PIR capacity  is
\begin{equation}
  \const{C}^{[n,k]}_f\eqdef\frac{n-k}{n}\inv{\left[1-\Bigl(\frac{k}{n}\Bigr)^f\right]},
  \label{eq:PIRcapacity}  
\end{equation}
where superscript “$[n,k]$” indicates the code parameters of the underlying MDS storage code. When the number of files
$f$ tends to infinity, \eqref{eq:PIRcapacity} reduces to
$\const{C}^{[n,k]}_\infty \eqdef  \lim_{f\to\infty}  \const{C}^{[n,k]}_f = \frac{n-k}{n}$,  
which we refer to as the asymptotic MDS-PIR capacity. Note that for the case of non-MDS linear codes, the PIR capacity
is unknown.




  \begin{table*}[!t]
    \centering
    \caption{Protocol~1 with a $[5,3]$ non-MDS-PIR capacity-achieving code for $f=2$}
    \label{tab:nonMDS-PIRcapacity-achieving-code_n5k3}
    \vspace{-7.5mm}
    \begin{IEEEeqnarray*}{rCl}
      \begin{IEEEeqnarraybox}[
        \IEEEeqnarraystrutmode
        \IEEEeqnarraystrutsizeadd{3pt}{1pt}]{v/c/V/c/v/c/v/c/v/c/v/c/v/c/v}
        \IEEEeqnarrayrulerow\\
        & && && \text{Node } 1  && \text{Node } 2 &&
        \text{Node } 3 && \text{Node } 4 &&
        \text{Node } 5 &\\
        \hline\hline
        & && && y^{(1)}_{2({\r 2}-1)+1,1} &&  y^{(1)}_{2({\r 1}-1)+1,2} && y^{(1)}_{2({\r 1}-1)+1,3} &&
        y^{(1)}_{2({\r 1}-1)+1,4} && y^{(1)}_{2({\r 1}-1)+1,5} &
        \\
        & && && y^{(1)}_{2({\r 2}-1)+2,1} &&  y^{(1)}_{2({\r 1}-1)+2,2} && y^{(1)}_{2({\r 1}-1)+2,3} &&
        y^{(1)}_{2({\r 1}-1)+2,4} && y^{(1)}_{2({\r 1}-1)+2,5} &
        \\*\cline{5-15}
        & && \rot{\rlap{\text{round} 1}}&&y^{(2)}_{3\cdot 0+{\r 2},1} &&  y^{(2)}_{3\cdot 0+{\r 1},2}
        && y^{(2)}_{5\cdot 0+{\r 1},3}  &&y^{(2)}_{3\cdot 0+{\r 1},4} && y^{(2)}_{3\cdot 0+{\r 1},5} &
        \\
        & \rot{\rlap{\text{repetition} 1}}&& &&y^{(2)}_{3\cdot 0+{\r 3},1} && y^{(2)}_{3\cdot 0+{\r 3},2}
        && y^{(2)}_{3\cdot 0+{\r 3},3}  &&y^{(2)}_{3\cdot 0+{\r 2},4} && y^{(2)}_{3\cdot 0+{\r 2},5}
        \\*\cline{3-15}
        & && \text{rnd}.~2 &&y^{(1)}_{2\cdot 3+{\r 2},1}+y^{(2)}_{3\cdot 0+{\b 1},1} &&
        y^{(1)}_{2\cdot 3+{\r 1},2}+y^{(2)}_{3\cdot 0+{\b 2},2}
        && y^{(1)}_{2\cdot 3+{\r 1},3}+y^{(2)}_{3\cdot 0+{\b 2},3} &&
        y^{(1)}_{2\cdot 3+{\r 1},4}+y^{(2)}_{3\cdot 0+{\b 3},4}&& y^{(1)}_{2\cdot 3+{\r 1},5}+y^{(2)}_{3\cdot 0+{\b 3},5} &
        \\*\hline\hline
        & && && y^{(1)}_{2({\r 3}-1)+1,1} &&  y^{(1)}_{2({\r 3}-1)+1,2} && y^{(1)}_{2({\r 3}-1)+1,3} &&
        y^{(1)}_{2({\r 2}-1)+1,4} && y^{(1)}_{2({\r 2}-1)+1,5} &
        \\
        & && && y^{(1)}_{2({\r 3}-1)+2,1} &&  y^{(1)}_{2({\r 3}-1)+2,2} && y^{(1)}_{2({\r 3}-1)+2,3} &&
        y^{(1)}_{2({\r 2}-1)+2,4} && y^{(1)}_{2({\r 2}-1)+2,5} &
        \\*\cline{5-15}
        & && \rot{\rlap{\text{round} 1}}&&y^{(2)}_{3\cdot 1+{\r 2},1} &&  y^{(2)}_{3\cdot 1+{\r 1},2}
        && y^{(2)}_{3\cdot 1+{\r 1},3}  &&y^{(2)}_{3\cdot 1+{\r 1},4} && y^{(2)}_{3\cdot 1+{\r 1},5} &
        \\
        & \rot{\rlap{\text{repetition} 2}}&& &&y^{(2)}_{3\cdot 1+{\r 3},1} && y^{(2)}_{3\cdot 1+{\r 3},2}
        && y^{(2)}_{3\cdot 1+{\r 3},3}  &&y^{(2)}_{3\cdot 1+{\r 2},4} && y^{(2)}_{3\cdot 1+{\r 2},5}
        \\*\cline{3-15}
        & && \text{rnd}.~2 &&y^{(1)}_{2\cdot 3+{\r 3},1}+y^{(2)}_{3\cdot 1+{\b 1},1} &&
        y^{(1)}_{2\cdot 3+{\r 3},2}+y^{(2)}_{3\cdot 1+{\b 2},2}
        && y^{(1)}_{2\cdot 3+{\r 3},3}+y^{(2)}_{3\cdot 1+{\b 2},3} &&
        y^{(1)}_{2\cdot 3+{\r 2},4}+y^{(2)}_{3\cdot 1+{\b 3},4}&& y^{(1)}_{2\cdot 3+{\r 2},5}+y^{(2)}_{3\cdot 1+{\b 3},5} &
        \\*\IEEEeqnarrayrulerow
      \end{IEEEeqnarraybox}
    \end{IEEEeqnarray*}
    \vskip -4ex 
  \end{table*} 
  
\subsection{MDS-PIR Capacity-Achieving Codes}
\label{sec:PIRachievable-rates_Symmetric}

In \cite{KumarLinRosnesGraellAmat17_1sub}, two symmetric PIR protocols for coded DSSs, named Protocol~1 and Protocol~2,
were proposed. 
Their PIR rates
depend on the following property of the underlying storage code $\code{C}$.
\begin{definition}
  \label{def:PIRachievable-rate-matrix}
  Let $\code{C}$ be an arbitrary $[n,k]$ code. A $\nu\times n$ binary matrix $\mat{\Lambda}_{\kappa,\nu}(\code{C})$ is
  said to be a \emph{PIR achievable rate matrix} for $\code{C}$ if the following conditions are satisfied.
  \begin{enumerate}
  \item \label{item:1} The Hamming weight of each column of $\mat{\Lambda}_{\kappa,\nu}$ is $\kappa$, and
  \item \label{item:2} for each matrix row $\vect{\lambda}_i$, $i\in\Nat{\nu}$, $\chi(\vect{\lambda}_i)$ always contains
    an information set.
  \end{enumerate}
\end{definition}

The following theorem gives the achievable PIR rate of Protocol~1 from \cite[Thm.~1]{KumarLinRosnesGraellAmat17_1sub}.
\begin{theorem}
  \label{thm:PIRachievable-rates_Symmetric}
  Consider a DSS that uses an $[n,k]$ code $\code{C}$ to store $f$ files. If a PIR achievable rate matrix
  $\mat{\Lambda}_{\kappa,\nu}(\code{C})$ exists, then the PIR rate
  \begin{IEEEeqnarray}{rCl}
    \const{R}_{f,\,\mathsf{S}}(\code{C})& \eqdef &
    \frac{(\nu-\kappa)k}{\kappa n}\inv{\left[1-\Bigl(\frac{\kappa}{\nu}\Bigr)^f\right]}
    \label{eq:finitePIRachievable-rate_symmetric}
  \end{IEEEeqnarray}
  is achievable. 
\end{theorem}

In \eqref{eq:finitePIRachievable-rate_symmetric}, we use subscript $\mathsf S$ to indicate that this PIR rate is
achievable by the symmetric Protocol~1 in \cite{KumarLinRosnesGraellAmat17_1sub}. Define
$\const{R}_{\infty,\,\mathsf{S}}(\code{C})$ as the limit of $\const{R}_{f,\,\mathsf{S}}(\code{C})$ as the number of
files $f$ tends to infinity, i.e.,
$\const{R}_{\infty,\,\mathsf{S}}(\code{C})\eqdef \lim_{f\to\infty} \const{R}_{f,\,\mathsf{S}}(\code{C})
=\frac{(\nu-\kappa)k}{\kappa n}$. The asymptotic PIR rate $\const{R}_{\infty,\,\mathsf{S}}(\code{C})$ is also achieved
by the file-independent Protocol~2 from \cite{KumarLinRosnesGraellAmat17_1sub}.


\begin{corollary}
  \label{cor:MDS-PIRcapacity-achieving-matrix}
  If a PIR achievable rate matrix $\mat{\Lambda}_{\kappa,\nu}(\code{C})$ with $\frac{\kappa}{\nu}=\frac{k}{n}$ exists
  for an $[n,k]$ code $\code{C}$, then the MDS-PIR capacity in \eqref{eq:PIRcapacity} is achievable.
\end{corollary}

\begin{definition}
  \label{def:MDS-PIRcapacity-achieving-codes}
  A PIR achievable rate matrix $\mat{\Lambda}_{\kappa,\nu}(\code{C})$ with $\frac{\kappa}{\nu}=\frac{k}{n}$ for an
  $[n,k]$ code $\code{C}$ is called an \emph{MDS-PIR capacity-achieving} matrix, and $\code{C}$ is referred to as an
  \emph{MDS-PIR capacity-achieving} code.
\end{definition}

%
The following theorem from \cite[Thm.~3]{KumarLinRosnesGraellAmat17_1sub} provides a necessary condition for the
existence of an MDS-PIR capacity-achieving matrix.
\begin{theorem}
  \label{thm:general-d_PIRcapacity-achieving-codes}
  If an MDS-PIR capacity-achieving matrix exists for an $[n,k]$ code $\code{C}$, then
  $d_s^{\code{C}}\geq\frac{n}{k}s$, $\forall\,s\in\Nat{k}$.
\end{theorem}


\section{PIR Capacity for MDS-PIR Capacity-Achieving Codes}
\label{sec:PIRcapacity_MDS-PIRcapacity-achieving-odes}


In this section, we prove that the PIR capacity of MDS-PIR capacity-achieving codes is equal to the MDS-PIR capacity.


\begin{theorem}
  \label{thm:converse_MDS-PIRcapacity-achieving-codes}
  Consider a DSS that uses an $[n,k]$ MDS-PIR capacity-achieving code $\code{C}$ to store $f$ files. Then, the maximum
  achievable PIR rate over all possible PIR protocols, i.e., the PIR capacity, is equal to the MDS-PIR capacity
  $\const{C}^{[n,k]}_{f}$ in \eqref{eq:PIRcapacity}.
\end{theorem}
\begin{IEEEproof}
    See \cite[App.~A]{lin18_1}.
\end{IEEEproof}


\Cref{thm:converse_MDS-PIRcapacity-achieving-codes} provides an expression for the PIR capacity for the family of
MDS-PIR capacity-achieving codes (i.e., \eqref{eq:PIRcapacity}). Moreover, for any finite number of files $f$ and in the
asymptotic case where $f$ tends to infinity, the PIR capacity can be achieved using Protocols~1 and 2 from
\cite{KumarLinRosnesGraellAmat17_1sub}, respectively.

\section{Asymmetry Helps: Improved PIR Protocols}
\label{sec:asymmetric-helps}

In this section, we present three asymmetric PIR protocols for non-MDS-PIR capacity-achieving codes, illustrating that
asymmetry helps to improve the PIR rate. By asymmetry we  mean that the number of symbols downloaded from the
different nodes is not the same, i.e., for any fixed $m\in\Nat{f}$, the entropies $\eHP{A_l^{(m)}}$, $l\in\Nat{n}$, may
be different. This is in contrast to the case of MDS codes, where any asymmetric protocol can be made symmetric while
preserving its PIR rate \cite{SunJafar17_1,SunJafar17_2,BanawanUlukus18_1}.
We start with a simple motivating example showing that the PIR rate of Protocol~1 from
\cite{KumarLinRosnesGraellAmat17_1sub} can be improved for some underlying storage codes.

\subsection{Protocol~1 From \cite{KumarLinRosnesGraellAmat17_1sub} is Not Optimal in General}
\label{sec:NotOptimal_symmetricPIR}

\begin{example}
  \label{ex:PIRrate_bad-code_n5k3}
  Consider the $[5,3]$ code $\code{C}$ with generator matrix
  \begin{equation*}
    \mat{G}=
    \begin{pmatrix}
      1 & 0 & 0 & 1 & 0
      \\
      0 & 1 & 0 & 1 & 0
      \\
      0 & 0 & 1 & 0 & 1
    \end{pmatrix}.
  \end{equation*}
  The smallest possible value of $\frac{\kappa}{\nu}$ for
  which a PIR achievable rate matrix exists is $\frac{2}{3}$ and a corresponding PIR achievable rate matrix is 
   \begin{equation*} 
    \mat{\Lambda}_{2,3}=
    \begin{pmatrix}
      0 & 1 & 1 & 1 & 1
      \\
      1 & 0 & 0 & 1 & 1
      \\
      1 & 1 & 1 & 0 & 0
    \end{pmatrix}.
  \end{equation*}
  It is easy to verify that $\mat{\Lambda}_{2,3}$ above is a PIR achievable rate matrix for code $\code{C}$. Thus, the
  largest PIR rate for $f=2$ files with Protocol~1 from \cite{KumarLinRosnesGraellAmat17_1sub} is
  $\const{R}_{2,\,\mathsf{S}}=\frac{3^3}{5\cdot 10}=\frac{27}{50}$. In
  Table~\ref{tab:nonMDS-PIRcapacity-achieving-code_n5k3} (taken from \cite[Sec.~IV]{KumarLinRosnesGraellAmat17_1sub}),
  we list the downloaded sums of code symbols when retrieving file $\mat{X}^{(1)}$ and $f=2$ files are stored. In the
  table, for each $m\in\Nat{2}$ and $\beta=\nu^f=3^2$, the interleaved code array $\mat{Y}^{(m)}$ with row vectors
  $\vect{y}^{(m)}_i=\vect{c}^{(m)}_{\pi(i)}$, $i\in\Nat{3^2}$, is generated (according to Protocol~1 from
  \cite{KumarLinRosnesGraellAmat17_1sub}) by a randomly selected permutation function $\pi(\cdot)$.

  Observe that since $\{2,3,4\}\subset\chi(\vect{\lambda_1})=\{2,3,4,5\}$ is an information set of $\code{C}$, the five
  sums of
  \begin{IEEEeqnarray*}{c}
    \Scale[0.95]{\bigl\{y^{(1)}_{2({\r 1}-1)+1,5}, y^{(1)}_{2({\r 1}-1)+2,5}, y^{(2)}_{3\cdot 0+{\r 1},5},
      y^{(1)}_{2\cdot 3+{\r 1},5}+y^{(2)}_{3\cdot 0+{\b 3},5}, y^{(2)}_{3\cdot 1+{\r 1},5}\bigr\}}
  \end{IEEEeqnarray*}
  are not necessarily required to recover $\mat{X}^{(1)}$.
  For privacy concerns, notice that the remaining sums of code symbols from the $5$-th node would be
  \begin{IEEEeqnarray*}{c}
    \Scale[0.95]{\bigl\{y^{(2)}_{3\cdot 0+{\r 2},5}, y^{(1)}_{2({\r 2}-1)+1,5}, y^{(1)}_{2\cdot({\r 2}-1)+2,5},
      y^{(2)}_{3\cdot 1+{\r 2},5}, y^{(1)}_{2\cdot 3+{\r 2},5}+y^{(2)}_{3\cdot 1+{\b 3},5}\bigr\}}.
  \end{IEEEeqnarray*}
 This ensures the privacy condition, since for every combination of files, the user downloads the same number of linear
  sums. This shows that by allowing asymmetry in the responses from the storage nodes, the PIR rate can be improved to
  $\frac{27}{50-5}=\frac{27}{45}=\frac{3}{5}$, which is much closer to the MDS-PIR capacity
  $ \const{C}^{[5,3]}_2=\frac{1}{1+\frac{3}{5}}=\frac{5}{8}$.
\end{example}

Example~\ref{ex:PIRrate_bad-code_n5k3} indicates that for a coded DSS using a non-MDS-PIR capacity-achieving code, there
may exist an asymmetric PIR scheme that improves the PIR rate of the symmetric Protocol~1 from
\cite{KumarLinRosnesGraellAmat17_1sub}.

\subsection{Protocol~A: A General Asymmetric PIR Protocol}
\label{sec:PIRachievable-rates_Asymmetric}

In this subsection, we show that for non-MDS-PIR capacity-achieving codes, by discarding the redundant coordinates that
are not required to form an information set within $\chi(\vect{\lambda}_i)$,
$i\in\Nat{\nu}$, it is always possible to obtain a larger PIR rate compared to that of Protocol~1 from
\cite{KumarLinRosnesGraellAmat17_1sub}.

\begin{theorem}
  \label{thm:PIRachievable-rates_Asymmetric}
  Consider a DSS that uses an $[n,k]$  code $\code{C}$ to store $f$ files. If a PIR achievable rate matrix 
  $\mat{\Lambda}_{\kappa,\nu}(\code{C})$ exists, then the PIR rate
  \begin{IEEEeqnarray}{rCl}
    \const{R}_{f,\,\mathsf{A}}(\code{C})& \eqdef &
    \Bigl(1-\frac{\kappa}{\nu}\Bigr)\inv{\left[1-\Bigl(\frac{\kappa}{\nu}\Bigr)^{f}\right]}
    \label{eq:RfA}
  \end{IEEEeqnarray}
  is achievable. 
\end{theorem}
\begin{IEEEproof}
  See \cite[App.~B]{lin18_1}.
\end{IEEEproof}


Proposition~\ref{prop:1} below can be easily verified using \cite[Lem.~2]{KumarLinRosnesGraellAmat17_1sub}.

\begin{proposition} \label{prop:1}
  Consider a DSS that uses an $[n,k]$ code $\code{C}$ to store $f$ files. Then,
  $\const{R}_{f,\,\mathsf{S}}(\code{C})\leq\const{R}_{f,\,\mathsf{A}}(\code{C})\leq\const{C}^{[n,k]}_f$ with equality
  if and only if $\code{C}$ is an MDS-PIR capacity-achieving code.
\end{proposition}


%

In the following, we refer to the asymmetric PIR protocol that achieves the PIR rate in
Theorem~\ref{thm:PIRachievable-rates_Asymmetric} as Protocol~A (thus the subscript $\mathsf A$ in
$\const{R}_{f,\,\mathsf{A}}(\code{C})$ in \eqref{eq:RfA}). Similar to Theorem~\ref{thm:PIRachievable-rates_Symmetric},
there also exists an asymmetric file-independent PIR protocol that achieves the asymptotic PIR rate
$\const{R}_{\infty,\,\mathsf{A}}(\code{C})\eqdef\lim_{f \to \infty}\const{R}_{f,\,\mathsf{A}}(\code{C}) =
1-\frac{\kappa}{\nu}$ and we simply refer to this protocol as the file-independent Protocol~A.\footnote{As for
  Protocol~1 and Protocol~2 from \cite[Remark~2]{KumarLinRosnesGraellAmat17_1sub}, $\mat{\Lambda}_{\kappa,\nu}(\code{C})$ can
be used for both Protocol~A and the file-independent Protocol~A.} 

\subsection{Protocol~B: An Asymmetric PIR Protocol for a Special Class of Non-MDS-PIR Capacity-Achieving Codes}
\label{sec:PIRachievable-rates_DirectSumMPcodes}

In this subsection, we focus on designing an asymmetric PIR protocol, referred to as Protocol~B, for a special class of
$[n,k]$ non-MDS-PIR capacity-achieving codes, where the code is isometric to a \emph{direct sum} of $P\in\Nat{n}$
MDS-PIR capacity-achieving codes \cite[Ch.~2]{macwilliamssloane77_1}. Without loss of generality, we assume that the
generator matrix $\mat{G}$ of an $[n,k]$ non-MDS-PIR capacity-achieving code $\code{C}$ has the structure
\begin{IEEEeqnarray}{rCl}
  \mat{G}& = &
  \begin{pmatrix}
    \mat{G}_1 &           &       &
    \\
              & \mat{G}_2 &       &
    \\
              &           &\ddots & 
    \\
              &           &       &\mat{G}_P
   \end{pmatrix},
   \label{eq:DirectSumCode-assumptions}
\end{IEEEeqnarray}
where $\mat{G}_p$, of size $k_p\times n_p$, is the generator matrix of a punctured MDS-PIR capacity-achieving
subcode $\code{C}^{\mat{G}_p}$, $p\in\Nat{P}$. 

\begin{theorem}
  \label{thm:PIRachievable-rates_DirectSumMPCodes}
  Consider a DSS that uses an $[n,k]$ non-MDS-PIR capacity-achieving code $\code{C}$ to store $f$ files. If the code
  $\code{C}$ is isometric to a direct sum of $P\in\Nat{n}$ MDS-PIR capacity-achieving codes as in
  \eqref{eq:DirectSumCode-assumptions}, then the PIR rate
  \begin{IEEEeqnarray*}{rCl}
    \const{R}_{f,\,\mathsf{B}}(\code{C})& \eqdef & \inv{\left(\sum_{p=1}^P
        \frac{k_p}{k}\inv{\Bigl(\const{C}_{f}^{[n_p,k_p]}\Bigr)}\right)}
  \end{IEEEeqnarray*}
  is achievable. Moreover, the asymptotic PIR rate
  \begin{IEEEeqnarray*}{c}
    \const{R}_{\infty,\,\mathsf{B}}(\code{C})\eqdef\lim_{f\to\infty}\const{R}_{f,\,\mathsf{B}}(\code{C}) =
    \inv{\left(\sum_{p=1}^P\frac{k_p}{k}\inv{\Bigl(\const{C}_{\infty}^{[n_p,k_p]}\Bigr)}\right)}
  \end{IEEEeqnarray*}
  is achievable by a file-independent PIR protocol.
\end{theorem}

\begin{IEEEproof}
  See \cite[App.~C]{lin18_1}.
\end{IEEEproof}

\begin{table*}[!t]
     \caption{Responses by Protocol~C with a $[9,5]$ non-MDS-PIR capacity-achieving code}
    \label{tab:nonMDS-PIRcapacity-achieving-code_n9k5}
    \centering
    \def\Hline{\noalign{\hrule height 2\arrayrulewidth}}
    \vskip -2.0ex 
    \begin{tabular}{@{}lccccccccc@{}}
      \Hline \\ [-2.0ex]
       \text{Subresponses}
        &\text{Node } 1 & \text{Node } 2 & \text{Node } 3 & \text{Node } 4 & \text{Node } 5
        & \text{Node } 6& \text{Node } 7 & \text{Node } 8& \text{Node } 9 \\[0.5ex]
      \hline
      \\ [-2.0ex] \hline  \\ [-2.0ex]
      \text{Subresponse} 1 & $I_1+x^{(m)}_{1,1}$ & $I_2$ & $I_3+x^{(m)}_{1,3}$ & $I_4+x^{(m)}_{1,4}$ & $I_5+x^{(m)}_{1,5}$
        & $I_4+I_5$ & $I_3+I_5$ & $I_3+I_4+I_5$ & $I_1+I_2+I_4+I_5$\\
        \text{Subresponse} 2 & $I_6$ & $I_7+x^{(m)}_{1,2}$ & & $I_9$ & $I_{10}$ & & & & $I_6+I_7+I_9+I_{10}$
    \end{tabular}
    \vskip -2ex 
\end{table*}

We remark that Protocol~B requires $\beta=\mathsf{LCM}(\beta_1,\ldots,\beta_P)$ stripes, where $\beta_p$, $p\in\Nat{P}$,
is the smallest number of stripes of either Protocol~1 or Protocol~2 for a DSS that uses only the punctured MDS-PIR
capacity-achieving subcode $\code{C}^{\mat{G}_p}$ to store $f$ files  \cite[App.~C]{lin18_1}.  

Theorem~\ref{thm:PIRachievable-rates_DirectSumMPCodes} can be used to obtain a larger PIR rate for the non-MDS-PIR
capacity-achieving code in Example~\ref{ex:PIRrate_bad-code_n5k3}.

\begin{example} \label{ex:ImprovedPIRrate_bad-code-n5k3}
  Continuing with~\cref{ex:PIRrate_bad-code_n5k3}, by elementary matrix operations, the generator matrix of the
  $[5,3]$ code of \cref{ex:PIRrate_bad-code_n5k3} is equivalent to the generator matrix
  \begin{IEEEeqnarray*}{rCl}
    \begin{pmatrix}
      1 & 0 & 1 & 0 & 0
      \\
      0 & 1 & 1 & 0 & 0
      \\
      0 & 0 & 0 & 1 & 1
    \end{pmatrix}
    & = &
    \begin{pmatrix}
      \mat{G}_1 &
      \\
      & \mat{G}_2
    \end{pmatrix}.
  \end{IEEEeqnarray*}
  It can easily be verified that both $\code{C}^{\mat{G}_1}$ and $\code{C}^{\mat{G}_2}$ are MDS-PIR capacity-achieving
  codes. Hence, from Theorem~\ref{thm:PIRachievable-rates_DirectSumMPCodes}, the asymptotic PIR rate
    $\const{R}_{\infty,\,\mathsf{B}} = 
    \inv{\left(\frac{2}{3}\frac{1}{1-\frac{2}{3}}+\frac{1}{3}\frac{1}{1-\frac{1}{2}}\right)}=\frac{3}{8}$
  is achievable. The rate $\const{R}_{\infty,\,\mathsf{B}}=\frac{3}{8}$ is strictly larger than both
  $\const{R}_{\infty,\,\mathsf{S}}=\frac{3}{10}$ and $\const{R}_{\infty,\,\mathsf{A}}=\frac{1}{3}$. 

\end{example}

\subsection{Protocol~C: Code-Dependent Asymmetric PIR Protocol}
\label{sec:examples_code-dependent-asymmetricPIRprotocol}

In this subsection, we provide a code-dependent, but file-independent asymmetric PIR protocol for non-MDS-PIR
capacity-achieving codes that cannot be decomposed into a direct sum of MDS-PIR capacity-achieving codes as in
\eqref{eq:DirectSumCode-assumptions}. The protocol is tailor-made for each class of storage codes.  
The main principle of the protocol is to further reduce the number of downloaded symbols by looking at punctured MDS-PIR
capacity-achieving subcodes.
Compared to Protocol~A, which is simpler and allows for a closed-form expression for its PIR rate, Protocol~C 
gives larger PIR rates.

The file-independent Protocol~2 from \cite{KumarLinRosnesGraellAmat17_1sub} utilizes \emph{interference symbols}. An
interference symbol can be defined through a summation as \cite{KumarLinRosnesGraellAmat17_1sub}
\begin{IEEEeqnarray*}{c}
  I_{(h-1)k+h'}\eqdef\sum_{m=1}^f\sum_{j=(m-1)\beta+1}^{m\beta} u_{h,j}x^{(m)}_{j-(m-1)\beta,h'},
\end{IEEEeqnarray*}
where $h,h' \in \Nat{k}$ and the symbols $u_{h,j}$ are chosen independently and uniformly at random from the same field
as the code symbols.

\begin{example}
  \label{ex:ImprovedPIRrate_bad-code_n9k5}
  Consider a $[9,5]$ code $\code{C}$ with generator matrix
  \begin{IEEEeqnarray*}{rCl}
    \mat{G}& = &
    \begin{pmatrix}
      1 & 0 & 0 & 0 & 0 & 0 & 0 & 0 & 1
      \\
      0 & 1 & 0 & 0 & 0 & 0 & 0 & 0 & 1 
      \\
      0 & 0 & 1 & 0 & 0 & 0 & 1 & 1 & 0
      \\
      0 & 0 & 0 & 1 & 0 & 1 & 0 & 1 & 1
      \\
      0 & 0 & 0 & 0 & 1 & 1 & 1 & 1 & 1
    \end{pmatrix}.
  \end{IEEEeqnarray*}
  It has $d_2^{\code{C}}=3<\frac{9}{5}\cdot 2$, thus it is not MDS-PIR capacity-achieving (see
  \cref{thm:general-d_PIRcapacity-achieving-codes}). Note that this code cannot be decomposed into a direct sum of
  MDS-PIR capacity-achieving codes as in \eqref{eq:DirectSumCode-assumptions}. 

  The smallest $\frac{\kappa}{\nu}$ for which a PIR achievable rate matrix exists for
  this code is $\frac{2}{3}$, and a corresponding PIR achievable rate matrix is
  \begin{IEEEeqnarray*}{rCl}
    \mat{\Lambda}_{2,3}=
    \begin{pmatrix}
      0 & 1 & 0 & 0 & 0 & 1 & 1 & 1 & 1
      \\
      1 & 0 & 1 & 1 & 1 & 1 & 1 & 1 & 1
      \\
      1 & 1 & 1 & 1 & 1 & 0 & 0 & 0 & 0
    \end{pmatrix}.
  \end{IEEEeqnarray*}
  The idea of the file-independent Protocol~2 from \cite{KumarLinRosnesGraellAmat17_1sub} is to use the information sets
  $\set{I}_1=\{2,6,7,8,9\}$ and $\set{I}_2=\{1,3,4,5,9\}$ to recover the $\beta k = 1\cdot 5$ requested file symbols
  that are located in $\set{I}_3=\{1,2,3,4,5\}$. Specifically, we use the information set $\set{I}_1$ to reconstruct the
  required code symbols located in $\comp{\chi(\vect{\lambda}_1)}=\{1,3,4,5\}$ and
  $\set{I}_2\subseteq\chi(\vect{\lambda}_2)=\{1,3,4,5,6,7,8,9\}$ to reconstruct the required code symbol located in
  $\comp{\chi(\vect{\lambda}_2)}=\{2\}$. Since the code coordinates $\{1,2,4,5,9\}$ form an $[n',k']=[5,4]$ punctured
  MDS-PIR capacity-achieving subcode $\code{C}^{\mat{G}'}$ with generator matrix
  \begin{IEEEeqnarray*}{rCl}
    \mat{G}'& = &
    \begin{pmatrix}
      1 & 0 & 0 & 0 & 1
      \\
      0 & 1 & 0 & 0 & 1
      \\
      0 & 0 & 1 & 0 & 1
      \\
      0 & 0 & 0 & 1 & 1
    \end{pmatrix},
  \end{IEEEeqnarray*}
  it can be seen that the code coordinates $\{1,4,5,9\}$ are sufficient to correct the erasure located in
  $\comp{\chi(\vect{\lambda}_2)}$. Therefore, compared to Protocol~A, we can further reduce the required number of
  downloaded symbols. The responses from the nodes when retrieving file $\mat{X}^{(m)}$ are listed in
  Table~\ref{tab:nonMDS-PIRcapacity-achieving-code_n9k5}. The PIR rate of Protocol~C is then equal to
    $\const{R}_{\infty,\,\mathsf{C}}=\frac{1\cdot 5}{n+n'}=\frac{5}{14} < \frac{4}{9}=\const{C}^{[9,5]}_{\infty}$,
  which is strictly larger than $\const{R}_{\infty,\,\mathsf{A}}=\frac{1}{3}$. It can readily
  be seen from Table~\ref{tab:nonMDS-PIRcapacity-achieving-code_n9k5} that the privacy condition in (\ref{eq:privacy}) is
  ensured.

\setlength{\textfloatsep}{18.60004pt plus 2.39996pt minus 4.79993pt}
\begin{table}[!t]
     \caption{PIR rate for different codes and protocols}
    \label{tab:results}
    \centering
    \def\Hline{\noalign{\hrule height 2\arrayrulewidth}}
    \vskip -2.0ex 
    \begin{tabular}{@{}lcccccc@{}}
      \Hline \\ [-2.0ex]
      Code & $\frac{\kappa}{\nu}$ & $\const{R}_{\infty,\,\mathsf{S}}$ & $\const{R}_{\infty,\,\mathsf{A}}$ & $\const{R}_{\infty,\,\mathsf{B}}$
      & $\const{R}_{\infty,\,\mathsf{C}}$ & $\const{C}^{[n,k]}_{\infty}$ \\[0.5ex]
      \hline
      \\ [-2.0ex] \hline  \\ [-2.0ex]
      $\mathcal C_1:[5,3]$  & $2/3$ &$0.3$ & $0.3333$ &  $0.375$ & $0.375$ & $0.4$ \\
      $\mathcal C_2:[9,5]$   & $2/3$ &$0.2778$ & $0.3333$ &  $-$ & $0.3571$ & $0.4444$ \\ 
      $\mathcal C_3:[7,4]$      &   $3/5$                                   &$0.3810$ & $0.4$    &  $-$ & $0.4$    & $0.4286$ \\ 
      $\mathcal C_4:[11,6]$     &    $3/4$                                  &$0.1818$ & $0.25$   & $-$  & $0.2824$ & $0.4545$ \\ 
    \end{tabular}
    \vskip -2ex 
\end{table}

  Finally, we remark that, using the same principle as outlined above, other punctured MDS-PIR capacity-achieving
  subcodes can be used to construct a valid protocol, giving the same PIR rate.  For instance, we could pick the two
  punctured subcodes $\code{C}^{\mat{G}_1}$ and $\code{C}^{\mat{G}_2}$
  with generator matrices
  \begin{IEEEeqnarray*}{rCl}
    \mat{G}_1& = &
    \begin{pmatrix}
      1 & 0 & 0 & 0 & 1 & 1 
      \\
      0 & 1 & 0 & 1 & 0 & 1 
      \\
      0 & 0 & 1 & 1 & 1 & 1 
    \end{pmatrix}\;
    \textnormal{ and }\;
    \mat{G}_2=
    \begin{pmatrix}
      1 & 0 & 1 
      \\
      0 & 1 & 1 
    \end{pmatrix},
  \end{IEEEeqnarray*}
  respectively.
\end{example}

\Cref{ex:ImprovedPIRrate_bad-code_n9k5} above illustrates the main working principle of Protocol~C and how the redundant
set of code coordinates is taken into account. Its general description is given in \cite{lin18_1}.
However, some numerical results are given below, showing that it can attain larger PIR rates than Protocol~A.

\section{Numerical Results}
\label{sec:numerical-results}

In \cref{tab:results}, we compare the PIR rates for different protocols using several binary linear codes. The second
column gives the smallest fraction $\frac{\kappa}{\nu}$ for which a PIR achievable rate matrix exists. In the table,
code $\code{C}_1$ is from Example~\ref{ex:PIRrate_bad-code_n5k3}, code $\code{C}_2$ is from Example~\ref{ex:ImprovedPIRrate_bad-code_n9k5},
$\code{C}_3$ is a $[7,4]$ code with generator matrix $(1,2,4,8,8,14,5)$ (in decimal form, e.g., $\trans{(1,0,1,1)}$ is
represented by $13$) and $d_3^{\code{C}_3}=5<\frac{7}{4}\cdot 3$, and $\code{C}_4$ is an $[11,6]$ code with generator
matrix $(1,2,4,8,16,32,48,40,24,56,55)$ and $d_3^{\code{C}_4}=4<\frac{11}{6}\cdot 3$. Note that $\code{C}_2$,
$\code{C}_3$, and $\code{C}_4$ cannot be decomposed into a direct sum of MDS-PIR capacity-achieving codes as in
\eqref{eq:DirectSumCode-assumptions}. For all presented codes except $\code{C}_3$, Protocol~C achieves strictly larger
PIR rate than Protocol~A, although smaller than the MDS-PIR capacity.

\section{Conclusion}
\label{sec:conclusion}

We proved that the PIR capacity for MDS-PIR capacity-achieving codes is equal to the MDS-PIR
capacity for the case of noncolluding nodes, giving the first family of non-MDS codes for which the PIR capacity
is known. We also showed that allowing asymmetry in the responses from the storage nodes yields larger PIR
rates compared to symmetric protocols in the literature when the storage code is a non-MDS-PIR capacity-achieving
code. We proposed three asymmetric protocols and compared them in terms of PIR rate for different storage codes.
\bibliographystyle{IEEEtran}
\bibliography{./defshort1,./biblio1}

\end{document}